\documentclass[10pt,conference]{IEEEtran}
\usepackage{lipsum}
\usepackage[normalem]{ulem} 
\usepackage{amsmath}
\usepackage{amssymb}
\usepackage{subfig}
\usepackage{algorithmic}
\usepackage{graphicx}
\usepackage{subcaption}
\usepackage{textcomp}
\usepackage{soul}
\usepackage{xcolor}
\usepackage{enumitem}
\usepackage{multirow}
\usepackage{fancyhdr}
\usepackage{colortbl}
\usepackage{wrapfig}
\usepackage{multicol}
\usepackage{algorithm2e}
\RestyleAlgo{ruled}
\usepackage{pifont}%
\usepackage{makecell}
\usepackage{tikz}
\usepackage{float}
\usepackage{booktabs}
\usepackage{etoolbox}
\usepackage{listings}
\usepackage{hyperref}
\usepackage{cleveref}
\crefformat{section}{\S#2#1#3} %
\crefformat{subsection}{\S#2#1#3}
\crefformat{subsubsection}{\S#2#1#3}
\usepackage{url}
\usepackage[table]{xcolor} 
\usepackage{tikz} \usepackage{tikz} \usetikzlibrary{arrows.meta,shapes.geometric,decorations.pathreplacing}

\usepackage{xspace}
 
\newcommand{\SYSBASE}{OneDSE}          %
\newcommand{\MINDTAG}{MIND}            %
\newcommand{\SAILTAG}{SAIL}            %
 
\newcommand{\SYS}{\SYSBASE\xspace}                          %
\newcommand{\NAME}{\SYSBASE-\MINDTAG\xspace}                %
\newcommand{\NAMERL}{\SYSBASE-\MINDTAG{}$+$RL\xspace}       %
\newcommand{\NAMESEARCH}{\SYSBASE-\SAILTAG\xspace}          %

\newcommand{\NAMESEARCHRL}{\SYSBASE-\SAILTAG~(RL-seeded)\xspace}

\AtBeginDocument{%
  }

\newcommand{\hpcayear}{2027}

\definecolor{colorRevA}{rgb}{0.85, 0.90, 0.98}
\definecolor{colorRevB}{rgb}{0.97, 0.80, 0.796}
\definecolor{colorRevC}{rgb}{0.87, 0.83, 0.90}
\definecolor{colorRevD}{rgb}{0.83, 0.91, 0.83}
\definecolor{colorRevE}{rgb}{1, 0.90, 0.80}
\definecolor{colorRevF}{rgb}{0.89, 0.73, 0.89}
\definecolor{blond}{rgb}{0.98, 0.94, 0.75}

\begin{document}

\title{OneDSE: Metric-Conditioned Inverse Modeling and Active Search for Sample-Efficient DSE}

\author{
  \ifdefined\hpcacameraready
    \IEEEauthorblockN{\hpcaauthors{}}
      \IEEEauthorblockA{
        \hpcaaffiliation{} \\
        \hpcaemail{}
      }
  \else
    \IEEEauthorblockN{Ritik Raj\IEEEauthorrefmark{2},
Akshat Ramachandran\IEEEauthorrefmark{2},
Danny Samuel\IEEEauthorrefmark{3},
Jeff Nye\IEEEauthorrefmark{4},
Shashank Nemawarkar\IEEEauthorrefmark{4},
Tushar Krishna\IEEEauthorrefmark{2}}
\IEEEauthorblockA{\IEEEauthorrefmark{2}Georgia Institute of Technology}
\IEEEauthorblockA{\IEEEauthorrefmark{3}University of Michigan}
\IEEEauthorblockA{\IEEEauthorrefmark{4}Condor Computing}
  \fi 
}

\fancypagestyle{camerareadyfirstpage}{%
  \fancyhead{}
  \renewcommand{\headrulewidth}{0pt}
  \fancyhead[C]{
    \ifdefined\aeopen
    \parbox[][12mm][t]{13.5cm}{\hpcayear{} IEEE International Symposium on High-Performance Computer Architecture (HPCA)}    
    \else
      \ifdefined\aereviewed
      \parbox[][12mm][t]{13.5cm}{\hpcayear{} IEEE International Symposium on High-Performance Computer Architecture (HPCA)}
      \else
      \ifdefined\aereproduced
      \parbox[][12mm][t]{13.5cm}{\hpcayear{} IEEE International Symposium on High-Performance Computer Architecture (HPCA)}
      \else
      \parbox[][0mm][t]{13.5cm}{\hpcayear{} IEEE International Symposium on High-Performance Computer Architecture (HPCA)}
    \fi 
    \fi 
    \fi 
    \ifdefined\aeopen 
      \includegraphics[width=12mm,height=12mm]{ae-badges/open-research-objects.pdf}
    \fi 
    \ifdefined\aereviewed
      \includegraphics[width=12mm,height=12mm]{ae-badges/research-objects-reviewed.pdf}
    \fi 
    \ifdefined\aereproduced
      \includegraphics[width=12mm,height=12mm]{ae-badges/results-reproduced.pdf}
    \fi
  }
  \fancyfoot[C]{}
}
\fancyhead{}
\renewcommand{\headrulewidth}{0pt}

\maketitle

\ifdefined\hpcacameraready 
  \thispagestyle{camerareadyfirstpage}
  \pagestyle{empty}
\else
  \thispagestyle{plain}
  \pagestyle{plain}
\fi

\newcommand{\hpcaheight}{0mm}
\ifdefined\eaopen
\renewcommand{\hpcaheight}{12mm}
\fi

\begin{abstract}  

As Moore’s Law slows and power constraints tighten, architectural innovation becomes essential for performance gains, but its growing complexity expands modern high-performance processor design spaces.

We identify two key
challenges in prior CPU design space exploration (DSE) approaches: (a) \underline{short-horizon prediction is
forward-only}, modeling PPA (power, performance, area) metrics from design
parameters while designers start from metric targets, and
(b) \underline{long-horizon exploration is slow}, evaluating thousands of
candidates on cycle-accurate simulators. This work presents OneDSE, which unifies short-horizon design
prediction and long-horizon design optimization through
\textbf{M}etric-conditioned \textbf{IN}verse \textbf{D}esign
(\textbf{MIND}) and a \textbf{S}urrogate-\textbf{A}ssisted
\textbf{I}nverse \textbf{L}oop (\textbf{SAIL}).
\underline{First}, OneDSE-MIND inverts the prevailing recipe: conditioned on the workload,
it predicts the design that achieves target metrics, finding strong
design points in a handful of validations. An information-theoretic
analysis supports this inversion approach: the workload observation raises the
design information that metrics carry by 12--32\% otherwise a workload-blind approach makes richer metrics \emph{harder} to invert. \underline{Second}, OneDSE-SAIL embeds MIND in a
measurement loop in which fine-tuned
inverse proposals drive early sample efficiency while coordinated
multi-parameter operators secure the endpoints, under a distance-aware
acquisition. Results show that on five TailBench
workloads using gem5, MIND reaches, with as few as 1-58 validations,
design quality that ArchGym's genetic algorithm needs
$11\times$--$357\times$ as many evaluations to match (median
$68\times$). Further, SAIL attains a geometric-mean $0.98\times$ the full 6{,}400-evaluation
GA optimum with $12.5\times$ fewer online evaluations,  exceeding it
outright on one workload, versus $0.83\times$ for the strongest
budget-matched baseline (SMAC). Finally, we demonstrate generality beyond
CPUs by extending OneDSE to design space exploration of a DRAM memory
controller and the FEATHER reconfigurable AI accelerator.

\end{abstract}

\maketitle

\setcounter{section}{0}

\section{Introduction}
\label{sec:intro}

Since the Intel 4004~\cite{aspray1997intel}, the first commercial
microprocessor, CPUs have ridden a remarkable trajectory of
microarchitectural innovation, and they remain the universal compute
substrate: in embedded systems~\cite{fryer2005fpga, balfour2008energy},
data centers~\cite{talpes2022dojo, lotfi2012scale, tang2013optimizing},
avionics~\cite{keys2009advanced, george2018onboard},
HPC~\cite{vestias2014trends, sarood2013optimizing}, autonomous
driving~\cite{liu2017computer, lin2018architectural}, cryptographic
applications~\cite{gura2004comparing, kaneko2015cryptographic}, and,
increasingly, AI serving on general-purpose
cores~\cite{jeong2023vegeta, talpes2022dojo}. The agentic-AI era amplifies this role:
agent serving is orchestrated by, and heavily executes on, CPUs, whose
tool execution, retrieval, and coordination shape end-to-end latency
and throughput alongside GPU inference~\cite{raj2025cpu, chung2026characterizing}.

With transistor scaling at its physical
and economic limits~\cite{schaller1997moore, frank2001device,
razavieh2019challenges} and frequency scaling blocked by the power
wall~\cite{esmaeilzadeh2011dark, halpern2016mobile}, continued
performance gains rely on microarchitectural
innovation~\cite{halpern2016mobile}. Design space exploration (DSE)
is central to post-Moore microprocessor
design~\cite{li20222022, zheng2024bsse, bai2021boom}, and its search
space grows. Generations add parameters
(deeper cache and prefetch hierarchies, wider out-of-order engines,
richer branch prediction), while prior studies explored 20--30
parameters~\cite{ipek2006efficiently, bai2021boom, zheng2024bsse},
our out-of-order core exposes 64 parameters with 2--7
values each, ${\approx}10^{34}$ combinations.

Every prior CPU DSE framework searches the \emph{parameter space}~\cite{ipek2006efficiently, bai2021boom, zheng2024bsse}: it repeatedly proposes a configuration $\theta$ (cache geometry, pipeline widths, ROB sizing, and so on), asks a forward cost model $f\colon (\theta, w) \rightarrow m$ what metrics $m$ that configuration $\theta$ achieves on workload $w$, and iterates. This work turns the recipe around. A designer does not start from a configuration; they start from \emph{targets}, the performance and
area they need. \SYS therefore searches the \emph{metric space}: it
learns the inverse map $g\colon (m, w) \rightarrow \theta$ and answers
``which designs achieve these metrics on this workload?'' directly. The metric space is tiny, a handful of axes with small contiguous
ranges, so sweeping it is tractable where enumerating
$10^{34}$ configurations is not. The problem is that the inversion is a one-to-many function, that is, many design-workload pairs alias to the same metrics. Our key insight is that conditioning on the workload is what makes
inversion tractable, and we establish it quantitatively: across our
corpus, observing the workload raises the information that (IPC, area)
carries about the design by 12\%, and by 32\% once metrics include an
execution-trace summary (\S\ref{subsection:inverse_model}, Fig.~\ref{fig:mi}).

\begin{table*}
[t]\centering\small
\setlength{\tabcolsep}{3pt}
\resizebox{0.9\textwidth}{!}{%
\begin{tabular}{lllllc}
\toprule
Work & Search space & Cost model in loop & Workload aware &
\begin{tabular}[c]{@{}l@{}}One-shot\\ prediction\end{tabular} &
\begin{tabular}[c]{@{}c@{}}Gem5 Evals to\\ near-optimal\end{tabular} \\
\midrule
Exhaustive / ILP~\cite{padmanabhan2011optimal, niemann1997algorithm} &
  Parameter & Simulator & Per run & No & Intractable \\
Metaheuristics~\cite{chis2013multi, sengupta2012multi} &
  Parameter & Simulator & Per run & No & Thousands \\
ArchGym~\cite{krishnan2023archgym} &
  Parameter & Simulator & Per run & No & Thousands \\
BOOM-Explorer~\cite{bai2021boom} &
  Parameter & DKL-GP $+$ VLSI flow & No & No & ${>}512$ \\
BSSE~\cite{zheng2024bsse} &
  Parameter & Co-kNN predictor~\cite{zhou2007semisupervised} & No & No &
  ${>}512$ \\
GP-EI~\cite{jones1998efficient} &
  Parameter & GP surrogate (EI) & No & No & ${>}512^{\dagger}$ \\
SMAC~\cite{smac} / TuRBO~\cite{eriksson2019scalable} &
  Parameter & RF / trust-region GP & No & No & ${>}512^{\dagger}$ \\
\midrule
\textbf{\NAME{} (ours)} & \textbf{Metric} &
  \textbf{Trace transformer} &
  \textbf{Trace-conditioned} & \textbf{Yes ($K{\leq}58$)} &
  \textbf{1-58} \\
\textbf{\NAMESEARCH{} (ours)} & \textbf{Metric}\,$\to$\,\textbf{parameter} &
  \textbf{\NAME{} $+$ gem5} &
  \textbf{Trace-conditioned} & \textbf{Via \NAME{}} & \textbf{512} \\
\bottomrule
\end{tabular}}
\caption{Related CPU DSE works.
$^{\dagger}$Geomean best design at $512$ evaluations vs.\ the full-budget
ArchGym-GA reference (\S\ref{sec:eval:search}): SMAC $0.83\times$,
GP-EI $0.68\times$;
\NAMESEARCH{} $0.98\times$.}
\label{tab:related_works}
\vspace{-5mm}
\end{table*}

While past techniques have made notable strides, the prevailing
recipe fails in both halves: its forward orientation points
prediction the wrong way (Challenge~1), and its parameter-space
navigation leaves long-horizon search slow and barrier-bound
(Challenge~2).

\noindent
\textbf{Challenge 1: Prediction points the wrong way.}
\textit{(a)~Forward-only cost models.} Fast surrogates of the
simulator~\cite{li2022simnet, pandey2024tao} and microarchitecture-level
predictors~\cite{ipek2006efficiently, bai2021boom, zheng2024bsse}
accelerate the question ``what does $\theta$ achieve?'', but the
designer's question runs the other way, and answering it with a forward
model still requires wrapping a search loop around the simulator.
\textit{(b)~Workload-specific or workload-blind.} Existing predictors
are trained on a fixed workload set~\cite{zheng2024bsse, bai2021boom}
or marginalize the workload away; the former cannot transfer across diverse CPU deployments, from server to embedded to
AI-on-CPU~\cite{talpes2022dojo, jeong2023vegeta},
while the latter, as our analysis shows, makes the inverse problem statistically \emph{harder} as metrics get richer.

\noindent
\textbf{Challenge 2: Exploration is sample-starved.}
\textit{(a)~Slow simulation.} Cycle-accurate simulators
\cite{binkert2011gem5, kim2012macsim} price every search step in minutes (${\approx}212$\,s per gem5
evaluation in our setting), and exact
methods (exhaustive, ILP~\cite{niemann1997algorithm,
lukasiewycz2008efficient}, branch-and-bound~\cite{padmanabhan2011optimal})
are infeasible at a large design space of $10^{34}$ configurations.
\textit{(b)~Sample-hungry search.} Metaheuristics such as genetic
algorithms~\cite{holland1992genetic, katoch2021review} need thousands
of evaluations to converge~\cite{krishnan2023archgym}, Gaussian-process Bayesian
optimization degrades on 64-dimensional mixed discrete
spaces~\cite{smac, eriksson2019scalable}, and tree-ensemble surrogates' uncertainty
collapses exactly where exploration is needed
(\S\ref{sec:method:search}).
\textit{(c)~Coordinated barriers.} Better designs are often separated
from known ones by valleys, not slopes. Widening a core's front end,
for example, pays off only once the ROB, register files, and issue
queue scale up with it; widen the fetch alone and the extra
instructions simply stall on the unchanged back end, so the
intermediate design measures \emph{worse} than the one it came from. In our design space, high-fitness design
families sit a median of 32--37 coordinated parameter
changes away from the corpus elites (\S\ref{subsection:barrier}), while a useful mutation moves at most six parameters, so
every step along the only paths mutation can take is penalized. A
search that learns from feedback thus receives negative reinforcement
precisely along the routes that lead to better designs
(\S\ref{sec:eval:analysis}).

These challenges shape our contributions: OneDSE-MIND with reward-guided fine-tuning targets Challenge~1, while \NAMESEARCH{} targets Challenge~2. We contribute:

\begin{itemize}[leftmargin=10pt]
\item \textbf{\underline{\NAME}} (\textbf{M}etric-conditioned \textbf{IN}verse
\textbf{D}esign): a transformer over instruction-trace chunks that
inverts $(m, w) \rightarrow \theta$. A metric-target sweep then maps
a workload's reachable frontier in one shot. We also give the first
quantitative, information-theoretic demonstration that workload
conditioning makes CPU-design inversion tractable (\S\ref{subsection:inverse_model}).
 \item \textbf{\underline{Reward-guided fine-tuning}}: supervised inversion
 regresses toward the conditional mean of the dense corpus
 interior~\cite{bishop1994mixture, liu2018training, yang2021delving}.
 A REINFORCE~\cite{williams1992simple} objective against a corpus-distilled surrogate therefore
 pushes \NAME's predictions from that interior to the frontier
 (\S\ref{sec:method:rl}). The gain is a $+12\%$ geometric-mean best
 validated design within at most 121 gem5 validations, at negligible
 training cost.
 \item \textbf{\underline{\NAMESEARCH}} (\textbf{S}urrogate-\textbf{A}ssisted
\textbf{I}nverse \textbf{L}oop): a measurement loop with clear
role separation. \NAME's metric-targeted proposals, updated via
online fine-tuning, drive early-budget sample efficiency, while
coordinated multi-parameter operators cross barriers in
Challenge~2(c) and set the endpoints. A distance-aware acquisition
over a tree-ensemble surrogate selects each batch
(\S\ref{sec:method:search}).
\end{itemize}

We evaluate \SYS on five TailBench latency-critical
workloads~\cite{tailbench} in event-driven gem5 v25.1 with McPAT area models, against the ArchGym~\cite{krishnan2023archgym} reference search algorithms
under matched corpora and budgets (\S\ref{sec:eval}). 
Our results show that with as few as $K{=}1$--$58$ validations, \NAME reaches design quality that
ArchGym-GA needs $11\times$--$357\times$ as many evaluations to match
(median $68\times$) and that ArchGym-PPO never matches on four of five
workloads within its entire 6{,}400-evaluation budget. Closing the loop, \NAMESEARCH reaches a geometric-mean $0.98\times$ the
\emph{full} 6{,}400-evaluation ArchGym-GA optimum with only 512 online
evaluations, exceeding the performance on sphinx, while matching on
moses. At the same budget, SMAC reaches up to $0.83\times$, 
GP-EI Bayesian optimization up to $0.68\times$, and warm-started
ArchGym-GA up to $0.72\times$ at the same budget. 
The designs it finds are new, i.e., ArchGym does not evaluate any of these
configurations, and these designs cross coordinated
multi-parameter barriers that defeat mutation-based search. To support reproducibility and future CPU-DSE research, we will release the \SYSBASE{} implementation, gem5 configurations, and the design-metric-trace corpus generated for this study.

\section{Background}
\label{section:Background}

\subsection{Design Space Exploration}
\label{subsection:background_dse}

DSE search methods divide into \emph{exact} and \emph{non-exact}
families. Exact methods, exhaustive enumeration, integer linear
programming~\cite{niemann1997algorithm, lukasiewycz2008efficient}, and
branch-and-bound~\cite{padmanabhan2011optimal}, guarantee global
optimality by systematically partitioning the space and pruning regions
that provably cannot improve on the incumbent. Their worst-case cost is
exponential, which rules them out at the ${\approx}10^{34}$ scale of
our design space.

Non-exact methods~\cite{wang2006design, de2016design} trade optimality
guarantees for feasible search times using metaheuristics such as
genetic algorithms (GA)~\cite{holland1992genetic,
mitchell1998introduction, katoch2021review}, simulated annealing
(SA)~\cite{kirkpatrick1983optimization, van1987simulated,
bertsimas1993simulated}, and artificial bee colony (ABC)
optimization~\cite{karaboga2005idea, karaboga2007powerful,
gao2012global}. All share a two-step skeleton: a \emph{generation}
process that proposes new configurations (mutation and crossover in GA,
neighborhood perturbation in SA, employed bees in ABC) and a
\emph{selection} process that decides which configurations survive to
the next iteration, within one maintained configuration (SA) or a
population (GA, ABC). Because every generated configuration is priced
by the underlying simulator, these methods need thousands of
evaluations to converge on CPU design spaces~\cite{krishnan2023archgym}.
\NAMESEARCH{} keeps this generation-selection skeleton
(\S\ref{sec:method:search}) but replaces blind generation with
model-directed proposals and simulator-priced selection with
surrogate-priced selection.

\subsection{Bayesian Optimization}
\label{subsection:background_bo}

Bayesian optimization (BO)~\cite{jones1998efficient,
shahriari2015taking} is the framework for optimizing
expensive black-box functions under tight evaluation budgets, and it
supplies our strongest budget-matched baselines
(\S\ref{sec:eval:method}). BO maintains a probabilistic
\emph{surrogate} of the objective (classically a Gaussian process
(GP), whose posterior provides a mean prediction $\mu(x)$ and an
uncertainty $\sigma(x)$ per candidate) and an
\emph{acquisition function}, such as expected improvement
(EI)~\cite{jones1998efficient}, scoring candidates by combining
$\mu$ and $\sigma$ to balance exploiting promising regions against
exploring uncertain ones. Iterations evaluate the
acquisition-maximizing candidate (or a batch), update the surrogate,
and repeat.

Two properties strain classical BO. First, GP kernels
presume smooth, low-dimensional, continuous inputs; a 64-dimensional
space of discrete and categorical parameters degrades the model
and acquisition search, and posterior updates scale cubically
with observations~\cite{shahriari2015taking}. SMAC~\cite{smac}
replaces the GP with a random-forest surrogate that handles mixed
discrete spaces natively, and TuRBO~\cite{eriksson2019scalable} restores tractability
in high dimension by confining acquisition to adaptive trust regions.
Second, the tree ensembles that make mixed spaces tractable lose the
GP's calibrated uncertainty: outside the training support, all ensemble
members saturate toward the same prediction, so the spread $\sigma(x)$
\emph{collapses} exactly where uncertainty is largest.
\S\ref{sec:method:search} builds on the surrogate-guided skeleton
and compensates for this collapse with a distance-aware
acquisition, while adding the proposal families: metric-targeted
inverse predictions and coordinated operators that kernel-local
acquisition optimization does not generate.

\section{Motivation}
\label{section:Motivation}
 
\subsection{Metrics Are a Joint Product of Design and Workload}
\label{subsection:IPC Variation across workloads}

\begin{figure}
 \includegraphics[width=0.9\linewidth]{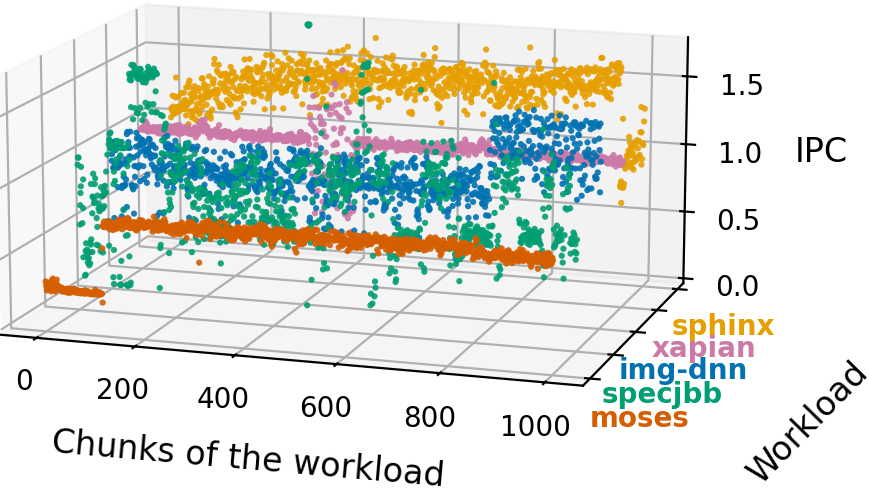}
 \caption{Chunk-level IPC across 1{,}000 execution-trace chunks of
  each of the five TailBench workloads.}
  \label{fig:motivation}
  \vspace{-5mm}
\end{figure}
 
Fig.~\ref{fig:motivation} plots per-chunk IPC over 1{,}000
execution-trace chunks of each TailBench workload. Two observations
motivate \SYS's design. First, the \emph{workload term dominates}:
sphinx sustains ${\approx}1.3$--$1.6$ IPC where moses sits near
$0.35$, a ${\approx}4\times$ spread produced by the workloads alone.
Any map between designs and metrics, forward or inverse, that
ignores the workload conflates these regimes, and no amount of model
capacity recovers what the input never contained. Second,
\emph{workload identity alone is not enough}: specjbb's chunks
disperse across nearly the full IPC range, and moses shifts regime
mid-run, so a single label per workload discards the phase structure
that determines which design resources matter when. This is why \NAME
conditions on the execution trace itself, consuming chunk-level
content rather than a workload ID, and why richer, trace-level metric
representations pay off, a dependence
\S\ref{subsection:inverse_model} pins down in bits
(Fig.~\ref{fig:mi}).
 
\subsection{The Design Space Cannot Be Brute-Forced}
\label{subsection:Inaccurate search}
 
The 64-parameter design
space constitutes ${\approx}10^{34}$ raw combinations. The scale forecloses exact
search regardless of how fast the cost model becomes. Grant a
hypothetical oracle that prices a design in $1\,\mu$s, eight orders of magnitude faster than gem5's $\approx212$ s and well below any learned surrogate's practical inference: exhausting the space would still
take ${\approx}3 \times 10^{20}$ years, and if that oracle were
$99.99\%$ accurate, it would misprice ${\approx}10^{30}$ designs along
the way, ample room to bury every true optimum. Metaheuristics are
the standard escape, but they trade away optimality
guarantees~\cite{yang2014metaheuristic, gandomi2013metaheuristic},
offer little convergence theory on spaces like
ours~\cite{yang2010nature, yang2010engineering}, and, crucially, still
price every generated configuration with the real simulator: the
6{,}400-evaluation ArchGym-GA reference campaign we compare against
costs roughly 700--900 minutes of wall-clock per workload on 32
workers (\S\ref{sec:eval}). Metaheuristics at
least walk the space step by step; the next subsection shows the steps
themselves are penalized.
 
\subsection{Good Designs Are Beyond Mutation Reach}
\label{subsection:barrier}
 
\begin{figure}
  \includegraphics[width=\linewidth]{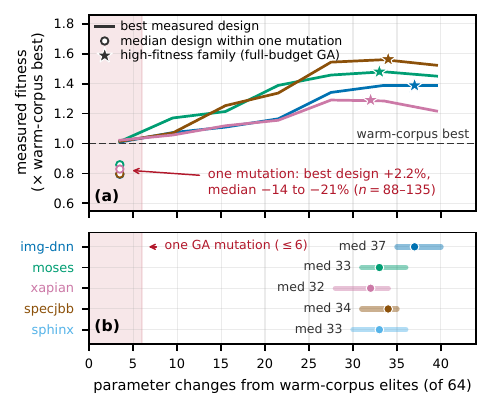}
  \caption{(a)~Best and median measured fitness vs.\ Hamming distance
  from the warm-corpus elite-128, normalized per workload to the
  corpus best; shaded: one GA mutation (${\leq}6$ changes).
  (b)~Distance from each full-budget-GA top-50 family to the corpus
  elites.}
  \label{fig:barrier}
  \vspace{-5mm}
\end{figure}
 
If the space cannot be enumerated, can local search at least walk
there? Fig.~\ref{fig:barrier}(a) pools every design the prior-method
campaigns measured and plots fitness against distance from the \emph{warm-corpus
elite-128}: the 128 highest-fitness designs in each workload's
offline corpus, the seed of every warm-started search and hence the
best designs known before a single online evaluation, making
$1.0\times$ the bar any search must clear. The one-mutation shell (${\leq}6$ parameter
changes) is well sampled, $n{=}88$--$135$ measured designs per
workload, and it is a trap: its best design improves on the corpus
best by at most $2.2\%$ ($1.010$--$1.022\times$), while its median
lands $14$--$21\%$ \emph{worse} ($0.79$--$0.86\times$), so a
mutation-based search sampling this shell overwhelmingly measures
regressions. Fitness climbs only tens of coordinated changes out: the
held-out full-budget-GA families peak near ${\approx}1.3$--$1.6\times$
the corpus best at median distances of $32$--$37$ changes
(Fig.~\ref{fig:barrier}(b)). On four of five workloads the elites are
themselves almost entirely ArchGym-GA designs (127/128 on img-dnn and
moses), so even the GA's best-known designs sit that far from the
families it eventually finds, and not one family member on any
workload lies within mutation reach (\S\ref{sec:eval:analysis}). Escaping requires proposals that move
many parameters at once and are aimed across the gap, which is
precisely what \SYS provides (\S\ref{sec:onedse}): \NAME's
metric-targeted inversion jumps to where the metrics say designs
should be, and \NAMESEARCH's coordinated operators cross the
intervening valleys.

\section{The \SYSBASE{} Framework}
\label{sec:onedse}
\label{sec:methodology}

\SYSBASE{} turns design-space exploration around: instead of predicting
the metrics of a proposed design, it predicts the \emph{design} that
achieves proposed metrics. This section presents the framework in four
parts: the data-generation pipeline that produces its training corpus
(\S\ref{subsection:data_gen}); \NAME{}, the workload-conditioned inverse
model, with a quantitative account of why workload conditioning makes
inversion tractable (\S\ref{subsection:inverse_model},
Figs.~\ref{fig:mind} and~\ref{fig:mi}); reward-guided fine-tuning, which pushes the model's
predictions from the dense corpus regions to its frontier
(\S\ref{sec:method:rl}); and \NAMESEARCH{}, which closes a measurement
loop around the model to reach within 2\% of full-search quality with
512 online evaluations against a 6{,}400-evaluation reference
(\S\ref{sec:method:search}, Fig.~\ref{fig:searchloop}).

\subsection{Data Generation}
\label{subsection:data_gen}
\label{subsection:data gen}

\textbf{Workloads and simulator.}
We target the five TailBench~\cite{tailbench} latency-critical
datacenter applications: img-dnn, moses, xapian, specjbb, and sphinx.
Each design is evaluated using the event-driven gem5 simulator~\cite{binkert2011gem5} with McPAT area models~\cite{li2009mcpat}. Event-driven gem5 is the community-standard cost model for CPU design
studies and captures the wrong-path and timing interactions that
trace-driven approximations miss. Each
evaluation reports the window-average IPC, a windowed IPC time series over the run, and the McPAT area. The design objective (\emph{fitness})
is $\mathrm{IPC}/\mathrm{area}\times10^{7}$.

\textbf{Design space.}
The search space is a 64-parameter out-of-order CPU core: superscalar widths, ROB/IQ/LSQ
sizing, physical register files, branch predictors, cache hierarchy,
prefetchers, clock, and memory configuration, each parameter taking
2--7 discrete or symbolic values, ${\approx}10^{34}$ raw combinations.
The technology node is itself a searched parameter ($\{22,32,45\}$\,nm),
entering through McPAT while core frequency is explored independently.
The sampler enforces microarchitectural validity by re-projection: the
integer register file must cover the ROB's in-flight destinations, and
pipeline widths obey commit $\leq$ issue $\leq$ writeback.

\textbf{Corpus.}
For each workload we simulate 1{,}000 uniformly random configurations.
This purely random corpus is the only data used to train the inverse
model offline. Each example pairs a design's parameters with its measured metrics
and the workload's execution-trace chunks, and the model is trained to
map (metrics, workload) back to parameters. \NAMESEARCH{} and every warm-started baseline later start from an offline corpus built on these samples (\S\ref{sec:eval:method}),
amortizing the one-time simulation cost across prediction, fine-tuning,
and search.

\subsection{\NAME{}: Workload-Conditioned Inverse Model}
\label{subsection:inverse_model}
\label{subsection: trace}
\label{section:OneDSE_metric}

\textbf{Architecture (Fig.~\ref{fig:mind}).}
\NAME{} encodes the workload as chunks of its dynamic instruction
trace ($s$ instructions each), tokenized by instruction type. This
keeps the dictionary and model compact, whereas encoding register indices or
immediates would inflate the dictionary by orders of magnitude.

Each token is embedded in $d$ dimensions, positional encodings supply
program-counter order, and $E_n$ encoder layers apply
\emph{instruction-window attention}: local
attention~\cite{beltagy2020longformer, parmar2018image} whose window
mirrors the out-of-order instruction window within which in-flight
instructions interact to determine latency, at $O(N)$ rather than
$O(N^2)$ cost. Each layer attends over a 100-instruction window; stacked over four
layers, the receptive field spans 400 instructions, exceeding the
largest reorder buffer in the design space (256 entries), so it
covers any two instructions that can be in flight together.
We use an encoder-only transformer~\cite{vaswani2017attention}: the
task is regression, and token-level autoregressive decoding over
millions of instructions is memory-bound and unnecessary. The encoded
chunk is mean-pooled into a single $d$-dimensional workload embedding,
concatenated with the metric vector $m$, and passed through $F_n$
feed-forward layers that predict all 64 parameters at once.

\begin{figure}[t]
 \centering
 \resizebox{\columnwidth}{!}{\definecolor{onedseC}{HTML}{D55E00}%
\definecolor{surrC}{HTML}{0072B2}%
\definecolor{gemC}{HTML}{009E73}%
\definecolor{embC}{HTML}{CC79A7}%
\definecolor{tokC}{HTML}{F0E442}%
\begin{tikzpicture}[x=1cm,y=1cm,
 font=\footnotesize\sffamily,
 >={Latex[length=2.2mm,width=1.8mm]},
 every path/.style={line width=0.8pt},
 box/.style={draw, rounded corners=2pt, align=center, inner sep=3.5pt,
   line width=0.9pt, text width=33mm},
 tracb/.style={box, draw=black!65, fill=tokC!45},
 embb/.style ={box, draw=embC!80!black, fill=embC!14},
 outb/.style ={box, draw=onedseC!85!black, fill=onedseC!12},
 sideb/.style={box, draw=embC!80!black, fill=embC!14, text width=23mm,
   font=\scriptsize\sffamily, anchor=west},
 metb/.style ={box, draw=gemC!75!black, fill=gemC!12, text width=23mm,
   font=\scriptsize\sffamily, anchor=west},
 dictb/.style={draw=surrC!80!black, fill=surrC!8, rounded corners=2pt,
   line width=0.9pt, inner sep=3pt, font=\scriptsize\sffamily,
   align=center},
 dim/.style  ={font=\scriptsize\itshape\sffamily, text=black!80,
   anchor=west, inner sep=1pt},
 opn/.style  ={circle, draw=black!75, fill=white, inner sep=0.6pt,
   font=\small, line width=0.9pt},
 note/.style ={font=\scriptsize\sffamily, align=center, inner sep=1.5pt},
 cell/.style ={draw=black!55, line width=0.35pt},
 ghost/.style={draw=none, fill=none, inner sep=0pt, outer sep=0pt}]

\node[tracb] (chunk) at (3.15,6.75)
 {\textbf{execution-trace chunk}\\ $s$ instructions};
\node[embb]  (emb)   at (3.15,4.60) {\textbf{token embeddings}\\ (dim $d$)};
\node[opn]   (plus)  at (3.15,3.85) {$\oplus$};
\node[embb]  (enc)   at (3.15,2.95)
 {$E_n$ \textbf{encoder layers}\\ instruction-window attention};
\node[embb, text width=27mm] (pool) at (3.15,1.55)
 {\textbf{workload embedding}};
\node[opn]   (cat)   at (3.15,0.80) {$\|$};
\node[outb]  (ffn)   at (3.15,-0.05)
 {$F_n$ \textbf{feed-forward layers} $+$ ReLU};
\node[outb, fill=onedseC!22] (out) at (3.15,-1.05)
 {\textbf{predicted design} $\theta$\\ all 64 parameters at once};

\node[dictb] (dict) at (0.30,5.35)
 {\textbf{dictionary}\\[0.4mm]
  \begin{tabular}{@{}l@{\ \ }r@{}}
   mov & 0\\ add & 1\\ cmp & 2\\
   \multicolumn{2}{@{}c@{}}{$\vdots$}\\
   jnz & $|V|{-}1$\\
  \end{tabular}};

\foreach \i/\t in {0/17,1/4,2/4,3/90,4/61,5/2}
  \node[cell, fill=tokC!45, minimum size=3.6mm, inner sep=0pt,
    font=\tiny] (tk\i) at (2.30+\i*0.40,5.40) {\t};
\node[font=\tiny, inner sep=1pt] (tdots) at (4.78,5.40) {$\cdots$};

\node[sideb] (pe)  at (5.75,3.85)
 {\textbf{positional encoding}\\ program-counter order};
\node[metb]  (met) at (5.75,0.80)
 {\textbf{metric targets} $m$\\ IPC, area, IPC-trace deciles};

\begin{scope}[shift={(5.85,3.05)}]
\foreach \r in {0,...,5} \foreach \c in {0,...,5}{
  \pgfmathtruncatemacro{\dd}{abs(\r-\c)}
  \ifnum\dd<2 \filldraw[cell, fill=embC!55]
    (\c*0.17,-\r*0.17) rectangle ++(0.15,0.15);
  \else \filldraw[cell, fill=black!6]
    (\c*0.17,-\r*0.17) rectangle ++(0.15,0.15);
  \fi}
\end{scope}
\node[ghost, minimum width=1.0cm, minimum height=1.0cm]
 (attn) at (6.35,2.70) {};
\node[note, anchor=west, align=left, text=black!85, text width=24mm]
 (lblAttn) at (5.85,1.80) {instruction window,\\ $O(N)$ attention};

\begin{scope}[shift={(5.75,-1.25)}]
\foreach \i/\v in {0/4,1/8,2/32,3/64}{
  \filldraw[cell, fill=onedseC!18] (\i*0.42,0) rectangle ++(0.38,0.32);
  \node[font=\tiny] at (\i*0.42+0.19,0.16) {\v};
  \node[font=\tiny, text=black!75] at (\i*0.42+0.19,0.50) {\i};}
\end{scope}
\node[ghost, minimum width=1.64cm, minimum height=0.62cm]
 (ordg) at (6.57,-0.94) {};
\node[note, anchor=west, align=left, text=black!85, text width=26mm]
 (lblOrd) at (5.75,-1.82) {ordinal ranks, rounded\\ to nearest (e.g.\ L2 kB)};

\node[dim] (dimA) at (5.10, 5.40) {$s{\times}1$};
\node[dim] (dimB) at (5.10, 4.60) {$s{\times}d$};
\node[dim] (dimC) at (5.10, 2.75) {$s{\times}d$};
\node[dim] (dimD) at (5.10, 1.55) {$1{\times}d$};
\node[dim] (dimE) at (5.10,-1.05) {$64$};

\draw[->] (chunk.south) -- (3.15,5.62);
\node[note, anchor=west, text=black!85] (lblTok)
 at (3.35,5.97) {tokenize by instruction type};
\draw[->] (dict.east) -- (2.10,5.40);
\draw[->] (3.15,5.18) -- (emb.north);
\draw[->] (emb.south)  -- (plus.north);
\draw[->] (pe.west)    -- (plus.east);
\draw[->] (plus.south) -- (enc.north);
\draw[->] (enc.south)  -- (pool.north);
\node[note, anchor=west, text=black!85] (lblPool) at (3.35,2.10)
 {mean-pool over $s$};
\draw[->] (pool.south) -- (cat.north);
\draw[->] (met.west)   -- (cat.east);
\draw[->] (cat.south)  -- (ffn.north);
\draw[->] (ffn.south)  -- (out.north);

\node[font=\footnotesize\sffamily, text=black!85, anchor=east]
 (gtag) at (1.20,-1.05) {$g\colon(m,w)\!\to\!\theta$};
\end{tikzpicture}}
 \caption{\textbf{\NAME{}}: trace chunks are tokenized by instruction
 type, embedded with program-counter positional encodings, encoded by
 $E_n$ instruction-window-attention layers, mean-pooled, concatenated
 with the targets $m$, and decoded by $F_n$ feed-forward layers into
 64 ordinal ranks.}
 \label{fig:mind}
 \vspace{-5mm}
\end{figure}

\textbf{Ordinal outputs and training.}
Parameters are predicted as \emph{ranks}: each parameter's discrete
value list is ordinally encoded (e.g., cache sizes $\{4,8,32,64\}$\,kB
become ranks $0$--$3$), keeping outputs in a comparable range where raw
values span six orders of magnitude; predictions round to the nearest
rank~\cite{williams2020ordinal}. Training
minimizes MSE with Adam (learning rate $10^{-3}$) over the per-workload
corpora of \S\ref{subsection:data_gen}.

\textbf{Why condition the inverse model on the workload.}
A metric vector $m$ is produced jointly by the design $\theta$ and the
workload $w$ through the forward map
$f\colon (\theta, w) \rightarrow m$. Prior predictive frameworks either
marginalize the workload away, collapsing distinct design--workload
pairs onto indistinguishable metrics and leaving the inverse
$g\colon m \rightarrow \theta$ under-determined, or fit disjoint
per-workload surrogates that stay well-posed but cannot pool design
information across workloads or transfer to unseen ones. \NAME{}
learns the forward structure once and inverts it \emph{jointly},
$g\colon (m, w) \rightarrow \theta$. By Fano's inequality, any inverse
predictor's error is bounded from below by the conditional entropy of
the design given its observations~\cite{cover2006elements}; conditioning on
$w$ lowers that entropy, raising the usable information from
$I(\theta; m)$ to
$I(\theta; [m, w]) = I(\theta; m) + I(\theta; w \mid m)$.

\begin{figure*}[t]
 \centering
 \includegraphics[width=\textwidth]{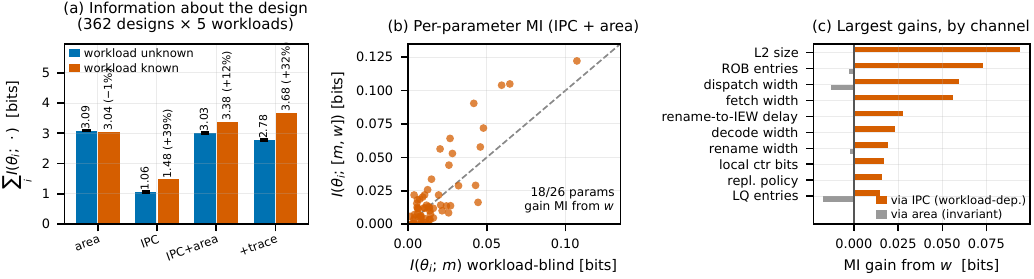}
 \caption{Mutual information (MI) between design parameters and
 metrics, workload unknown vs.\ known (362 designs $\times$ 5
 workloads; $k$NN estimator~\cite{ross2014mutual}).
 \textbf{(a)}~summed per-parameter MI, four metric representations;
 \textbf{(b)}~per-parameter MI for (IPC, area); \textbf{(c)}~ten
 largest gains by channel. Knowing the workload adds 12\% (IPC$+$area)
 and 32\% (trace-augmented).}
 \label{fig:mi}
 \vspace{-5mm}
\end{figure*}

Fig.~\ref{fig:mi} verifies that this gap is material on our design
space, over the 362 corpus designs simulated under all five workloads.
Conditioning on $w$ raises the summed per-parameter mutual information (MI) by 12\%
when $m$ comprises IPC and area, and by 32\% when $m$ adds the windowed-IPC decile summary the model consumes. Each observation
recovers a few bits of the $112.7$-bit summed design entropy (inversion
is genuinely hard), so the relative structure carries the signal, and
it is sharper than the totals. For a workload-blind observer the
extractable information \emph{shrinks} as workload-dependent
coordinates accumulate ($3.09 \to 3.03 \to 2.78$ bits from area-only to
trace-augmented, the trace step resolved at $5.2\sigma$ under paired
draws), because the observer cannot tell whether those coordinates
moved with the design or with the workload, so they blur rather than
sharpen the distinction between configurations. On the other hand, conditioning resolves
the ambiguity and reverses the ordering
($3.04 \to 3.38 \to 3.68$ bits), a
usable-information effect for a finite-sample observer~\cite{xu2020theory}, which is
exactly the regime a learned inverse model occupies. The channel decomposition
confirms the mechanism: McPAT computes area from $\theta$ alone,
bit-identical across workloads, and its channel moves by only $-0.05$
bits in aggregate ($|\Delta| \leq 0.017$ per parameter), a structural
null that doubles as an estimator control, while the gem5-produced IPC
channel gains 39\% (Fig.~\ref{fig:mi}a,c).

\textbf{Metric-target sweep.}
At deployment, \NAME{} answers a designer's question in one shot: given
target metrics, it emits the configuration predicted to achieve them.
To map a workload's reachable frontier we sweep a grid of 400
(IPC, fitness) targets, from corpus-typical to beyond the corpus best,
and validate the unique predictions in gem5 in descending-target order
as parallel batches. \S\ref{sec:eval:prediction} evaluates this sweep
against full search campaigns.

\subsection{Reward-Guided Fine-Tuning of the Inverse Model}
\label{sec:method:rl}

The inverse model is trained purely by supervised regression: it learns
the conditional expectation of designs given metrics, which makes it
accurate in dense regions of the corpus but conservative at the
frontier the designer actually cares about. We therefore fine-tune the
trained model with the reward-augmented objective
$\mathcal{L} = \mathcal{L}_{\mathrm{MSE}} -
\lambda\,\mathbb{E}\!\left[\mathrm{Perf}(\hat{x})\right]$,
which retains the supervised loss as an anchor while a performance term
pushes predictions toward high-fitness designs. Because a
cycle-accurate evaluation cannot sit inside a gradient loop,
$\mathrm{Perf}$ is a five-member gradient-boosted ensemble (differing
random seeds) fit on the training corpus (parameters $\rightarrow$ log
fitness), and its reward reaches the model through REINFORCE: Gaussian
exploration noise on the continuous ordinal outputs, a moving-average
baseline, and conditioning targets drawn half from corpus metrics and
half from the extrapolation region queried at deployment. Fine-tuning
updates only the prediction head on cached trace embeddings and costs
under 90 seconds per workload on a single GPU; $\lambda{=}0.05$
throughout.

Fig.~\ref{fig:prediction} shows the effect (dotted curves), measured
under the base sweep's accounting at a smaller validation budget (the
top-128 predicted designs per workload): fine-tuning raises the best
validated design on four of the five workloads, a geometric-mean
improvement of $12.2\%$ and up to $+23.2\%$ on specjbb, within
$K \leq 121$ gem5 evaluations (4--14 minutes on 32 workers). The
mechanism is knowledge extraction, not new search: the surrogate
distills the corpus, and the reward lets metric-conditioned prediction
reach the corpus frontier that supervised regression alone never emits;
on img-dnn the fine-tuned model predicts designs matching the
corpus-best fitness that the GA baseline required hundreds of
evaluations to find. Consistent with this mechanism, sphinx, whose offline corpus
(\S\ref{sec:eval:method}) contains only random designs, gives the
surrogate no frontier to distill, and fine-tuning there is neutral
($-2.2\%$), all at no change to the deployment interface.

\subsection{\NAMESEARCH{}: Closing the Loop Around Prediction}
\label{sec:method:search}

\NAME{} answers ``\emph{what design achieves these metrics?}'' in one
shot, yet that answer saturates below the optima long-running search
eventually finds: the model cannot extrapolate far beyond its training
distribution
(\S\ref{sec:eval:prediction}). \NAMESEARCH{} closes this gap by
embedding the inverse model in a measurement loop that repeatedly asks
a second question: ``\emph{given everything measured so far, which
designs are worth the next batch of simulations?}''

Formally, let $\mathcal{X}$ be the 64-parameter discrete design space,
$f(x)$ the fitness $\mathrm{IPC}(x)/\mathrm{area}(x)\times10^{7}$
obtained from one gem5$+$McPAT evaluation, and $\mathcal{D}_0$ the
offline corpus. \NAMESEARCH{} runs $R{=}16$ rounds; each round proposes
a candidate pool $\mathcal{C}$ ($|\mathcal{C}|{=}20{,}000$), selects a
batch $\mathcal{B}\subset\mathcal{C}$ ($|\mathcal{B}|{=}32$) for
simulation, and appends the measurements to the running dataset
$\mathcal{D}_r$, for a total online budget of 512 evaluations.
Fig.~\ref{fig:searchloop} overviews the loop: four components make it
effective, and a fifth, online fine-tuning of the inverse model, keeps
its proposals current.

\begin{figure*}[t]
 \centering
 \resizebox{\textwidth}{!}{\definecolor{onedseC}{HTML}{D55E00}
\definecolor{surrC}{HTML}{0072B2}
\definecolor{gemC}{HTML}{009E73}
\colorlet{pA}{onedseC!45}\colorlet{pB}{surrC!40}
\colorlet{pG}{black!22}\colorlet{pD}{gemC!45}
\begin{tikzpicture}[x=1cm,y=1cm,
 font=\footnotesize\sffamily,
 >={Latex[length=2.4mm,width=2mm]},
 every path/.style={line width=0.8pt},
 block/.style={draw, rounded corners=2.5pt, align=center, inner sep=4.5pt,
   text width=30mm, line width=0.9pt},
 prop/.style ={block, draw=onedseC!85!black, fill=onedseC!9,
   text width=42mm, align=left},
 scor/.style ={block, draw=surrC!80!black, fill=surrC!8},
 simb/.style ={block, draw=gemC!75!black, fill=gemC!10, text width=22mm},
 poolb/.style={block, draw=black!70, fill=black!5, text width=17mm},
 dbase/.style={draw=black!70, fill=black!5, cylinder,
   shape border rotate=90, aspect=0.12, minimum width=21mm,
   minimum height=11mm, align=center, inner sep=2pt, line width=0.9pt},
 group/.style={rounded corners=4pt, draw=black!35, fill=black!3,
   line width=0.7pt},
 stage/.style={font=\footnotesize\scshape, text=black},
 tagc/.style ={circle, draw=black!70, fill=white, inner sep=0.4pt,
   minimum size=3.4mm, font=\scriptsize\bfseries},
 panel/.style={rounded corners=4pt, draw=black!35, fill=black!2,
   line width=0.6pt},
 ptitle/.style={font=\footnotesize\bfseries, anchor=west, text=black},
 small/.style={font=\scriptsize\sffamily, text=black!90},
 lab/.style ={font=\scriptsize\sffamily, text=black, inner sep=1.5pt,
   fill=white},
 fb/.style ={->, draw=black!70},
 ft/.style ={->, draw=onedseC, dashed, line width=1.0pt},
 jdot/.style ={circle, fill=black!70, inner sep=0pt, minimum size=1.6pt},
 cell/.style ={draw=black!60, line width=0.35pt}]

\node[prop] (prop) at (2.55,9.30)
 {\textbf{Proposal sources}\\[0.3mm]
  \NAME{} proposals \textbf{(a)}\\ Coordinated operators \textbf{(b)}\\
  Local elite mutations\\ Random exploration};
\node[poolb] (pool) at (6.40,9.30)
 {candidate pool\\ $|\mathcal{C}|{=}20{,}000$};
\node[scor]  (surr) at (9.40,9.30)
 {\textbf{Surrogate ensemble}\\ fitness $\mu \pm \sigma$,\ \
  feasibility $p_{\mathrm{feas}}$};
\node[scor, text width=26mm] (sel) at (12.70,9.30)
 {\textbf{Batch selection} (c)\\ two scores $\to$ top 32};
\node[simb]  (sim) at (15.60,9.30)
 {\textbf{gem5 $+$ McPAT}\\ measure IPC, area};

\node[dbase] (dr) at (15.60,7.60) {measured set\\ $\mathcal{D}_r$};
\node[dbase] (dz) at (18.55,7.60) {offline corpus\\ $\mathcal{D}_0$};

\node[stage] (stProp)  at (2.55,10.50) {propose};
\node[stage] (stScore) at (9.40,10.50) {score};
\node[stage] (stSel)   at (12.70,10.50) {select};
\node[stage] (stSim)   at (15.60,10.50) {simulate};

\draw[->] (prop) -- (pool);
\draw[->] (pool) -- (surr);
\draw[->] (surr) -- (sel);
\draw[->] (sel)  -- (sim);
\draw[->] (sim.south) -- node[lab, right=0.4mm, pos=0.45] {append} (dr.north);
\draw[->] (dz.west) -- node[lab, above=0.5mm] {init} (dr.east);

\draw[fb] (dr.west) -- (9.40,7.60) -- (surr.south);
\node[lab] (lbRefit) at (11.90,7.60) {refit each round};

\node[jdot] at (10.60,7.60) {};
\draw[fb] (10.60,7.60) -- (10.60,6.95) -- (2.55,6.95) -- (prop.south);
\node[lab] (lbElite) at (6.30,6.95) {Pareto frontier $+$ elites};

\draw[ft] (dr.south) -- (15.60,6.45) -- (0.10,6.45) -- (0.10,9.30)
 -- (prop.west);
\node[lab, text=onedseC, font=\scriptsize\bfseries\sffamily] (lbFT)
 at (8.60,6.45) {fine-tune \NAME{} every 4 rounds ($+$ corpus replay)};

\begin{scope}[shift={(0.30,0.20)}]
\draw[panel] (0,0) rectangle (5.8,5.8);
\node[ptitle] (ttA) at (0.22,5.42) {(a)~Inverse proposals: aim at metrics};
\draw[->, black!70, line width=0.6pt] (0.75,1.35) -- (5.45,1.35);
\node[small] at (5.15,1.12) {area};
\draw[->, black!70, line width=0.6pt] (0.75,1.35) -- (0.75,4.90);
\node[small] at (1.05,4.75) {IPC};
\foreach \p in {(1.95,2.0),(2.45,2.5),(3.05,2.2),(2.75,3.0),(3.65,2.7),
 (4.15,3.3),(3.35,3.55),(4.65,2.9),(4.35,3.8),(2.2,3.2)}
 \fill[black!60] \p circle (1.3pt);
\node[small] at (4.95,2.25) {measured};
\draw[black!75, line width=0.8pt]
 (1.3,2.8) -- (1.8,3.4) -- (2.65,3.85) -- (3.8,4.2);
\foreach \p in {(1.3,2.8),(1.8,3.4),(2.65,3.85),(3.8,4.2)}
 \fill[black!85] \p circle (1.5pt);
\node[small] at (4.6,4.2) {frontier};
\foreach \p in {(2.25,3.7),(1.6,3.65),(3.25,4.15)}
 \node[star, star points=5, star point ratio=2.2, inner sep=0pt,
  minimum size=2.7mm, fill=onedseC, draw=onedseC!60!black] at \p {};
\foreach \p in {(1.35,4.3),(2.25,4.6)}
 \node[star, star points=5, star point ratio=2.2, inner sep=0pt,
  minimum size=2.7mm, fill=onedseC!55, draw=onedseC!60!black,
  line width=0.6pt] at \p {};
\node[star, star points=5, star point ratio=2.2, inner sep=0pt,
 minimum size=2.4mm, fill=onedseC!55, draw=onedseC!60!black,
 line width=0.6pt] at (2.86,4.95) {};
\node[small, anchor=west] at (3.02,4.95) {stretch ($\leq 1.5\times$ best)};
\node[small, align=center] at (2.9,0.62)
 {\NAME{} maps each target $(m^{\star}, w) \to \theta$;\\
  mutations around predictions seed the pool};
\end{scope}

\begin{scope}[shift={(6.65,0.20)}]
\draw[panel] (0,0) rectangle (5.8,5.8);
\node[ptitle] (ttB) at (0.22,5.42) {(b)~Coordinated operators};
\foreach \i in {0,...,9} \filldraw[cell, fill=pA]
 (0.45+\i*0.30,4.72) rectangle ++(0.26,0.26);
\foreach \i in {0,...,9} \filldraw[cell, fill=pB]
 (0.45+\i*0.30,4.32) rectangle ++(0.26,0.26);
\draw[->, black!70] (1.9,4.24) -- (1.9,3.98);
\foreach \i/\c in {0/pA,1/pA,2/pB,3/pA,4/pB,5/pB,6/pA,7/pB,8/pA,9/pB}
 \filldraw[cell, fill=\c] (0.45+\i*0.30,3.62) rectangle ++(0.26,0.26);
\node[small, anchor=west, align=left] at (3.62,4.30)
 {elite crossover:\\ mix two elites};
\foreach \i in {0,...,9} \filldraw[cell, fill=pG]
 (0.45+\i*0.30,2.92) rectangle ++(0.26,0.26);
\foreach \i in {0,...,9} \filldraw[cell, fill=pD]
 (0.45+\i*0.30,2.52) rectangle ++(0.26,0.26);
\draw[->, black!70] (1.9,2.44) -- (1.9,2.18);
\foreach \i/\c in {0/pG,1/pD,2/pG,3/pG,4/pD,5/pD,6/pG,7/pG,8/pD,9/pG}
 \filldraw[cell, fill=\c] (0.45+\i*0.30,1.82) rectangle ++(0.26,0.26);
\node[small, anchor=west, align=left] at (3.62,2.50)
 {path-relinking:\\ adopt 25--75\%\\ from a donor};
\foreach \i in {0,...,7} {\filldraw[cell, fill=pA]
 (0.45+\i*0.30,0.98) rectangle ++(0.26,0.26);
 \node[font=\tiny, text=black] at (0.58+\i*0.30,1.11) {$+$1};}
\foreach \i in {8,9} {\filldraw[cell, fill=pB]
 (0.45+\i*0.30,0.98) rectangle ++(0.26,0.26);
 \node[font=\tiny, text=black] at (0.58+\i*0.30,1.11) {$\uparrow$};}
\draw[decorate, decoration={brace, amplitude=2.5pt}, black!85]
 (0.45,1.34) -- (2.85,1.34)
 node[small, midway, above=2.5pt] {all 8 widths shift together};
\node[small, anchor=west, align=left] at (3.62,1.11)
 {width-group scaling\\ (back-end rescaled)};
\node[font=\scriptsize\itshape\sffamily, text=black!90, align=center,
 text width=50mm] at (2.9,0.45)
 {single-parameter steps score worse;\\ only joint moves cross barriers};
\end{scope}

\begin{scope}[shift={(13.00,0.20)}]
\draw[panel] (0,0) rectangle (5.8,5.8);
\node[ptitle] (ttC) at (0.22,5.42) {(c)~Batch selection};
\node[small, anchor=west] at (0.40,4.92)
 {exploit:\; $(\mu - 0.5\sigma)\; p_{\mathrm{feas}}$};
\node[small, anchor=west] at (0.40,4.47)
 {explore:\; $(\mu + \sigma)\; p_{\mathrm{feas}}
  \;+\; \lambda\, d(x)\, f^{\ast}$};
\node[small, align=center, text width=51mm] at (2.9,3.72)
 {$d(x)$ $=$ distance to measured data: tree-ensemble spread $\sigma$
  collapses off-support, so distance restores exploration};
\node[small] at (2.9,3.00) {one batch $=$ 32 slots};
\foreach \i in {0,...,15} \filldraw[cell, fill=onedseC!30]
 (0.35+\i*0.165,2.42) rectangle ++(0.145,0.30);
\foreach \i in {16,17} \filldraw[cell, fill=black!25]
 (0.35+\i*0.165,2.42) rectangle ++(0.145,0.30);
\foreach \i in {18,...,24} \filldraw[cell, fill=surrC!65]
 (0.35+\i*0.165,2.42) rectangle ++(0.145,0.30);
\foreach \i in {25,...,28} \filldraw[cell, fill=surrC!35]
 (0.35+\i*0.165,2.42) rectangle ++(0.145,0.30);
\foreach \i in {29,...,31} \filldraw[cell, fill=black!45]
 (0.35+\i*0.165,2.42) rectangle ++(0.145,0.30);
\foreach \x in {4,8,12} \draw[white, line width=0.7pt]
 (0.35+\x*0.165-0.01,2.42) -- ++(0,0.30);
\filldraw[cell, fill=onedseC!30] (0.40,1.88) rectangle ++(0.22,0.22);
\node[small, anchor=west] at (0.72,1.99) {family quotas: 4 each (16 slots)};
\filldraw[cell, fill=black!25] (0.40,1.53) rectangle ++(0.22,0.22);
\node[small, anchor=west] at (0.72,1.64) {random exploration (2 slots)};
\filldraw[cell, fill=surrC!65] (0.40,1.02) rectangle ++(0.22,0.22);
\node[small, anchor=west, align=left] at (0.72,1.02)
 {top-scored (14 slots):\\ $\tfrac12$ exploit $\cdot$
  $\tfrac14$ explore $\cdot$ rest diverse};
\node[font=\scriptsize\itshape\sffamily, text=black!90, align=center,
 text width=50mm] at (2.9,0.42)
 {quotas guarantee every family is measured; scores fill the rest};
\end{scope}
\end{tikzpicture}}
 \caption{\textbf{\NAMESEARCH{}}: each of 16 rounds draws a
 20{,}000-candidate pool (local, inverse, coordinated, random), ranks
 it with the surrogate ensemble under a distance-aware acquisition,
 and simulates the top 32; $\mathcal{D}_r$ refits the surrogate each
 round and fine-tunes \NAME{} every four rounds (dashed).
 \textbf{(a)--(c)}: the three distinguishing components.}
 \label{fig:searchloop}
 \vspace{-5mm}
\end{figure*}

\textbf{(1) Forward surrogate with decomposed targets.}
Each round, the loop refits a bootstrapped ensemble of five
gradient-boosted-tree regressors on $\mathcal{D}_r$. Each member
predicts $\log\mathrm{IPC}$ and $\log\mathrm{area}$ separately and
fitness is derived per member: area is near-deterministic in the
configuration while IPC carries the workload-dependent signal, so the
decomposition lets the ensemble spread expose disagreement in either
factor. A gradient-boosted feasibility classifier, trained on
success/failure outcomes once both classes exist, predicts whether a
candidate will complete in gem5 at all.

\textbf{(2) Metric-targeted inverse proposals.}
Each round, the loop queries \NAME{} at metric targets derived from the \emph{measured} Pareto frontier of $\mathcal{D}_r$:
the frontier points and interpolations between adjacent ones; their
$+5\%$-IPC and $-5\%$-area perturbations; and stretch targets up to
${\approx}1.5\times$ the best fitness observed in the offline corpus,
crossing an IPC grid with area anchors drawn from the measured elites'
area quantiles plus five fixed anchors ($8.9$--$9.75\,\mathrm{mm}^2$).
Mutations around the predicted configurations seed the candidate pool
with designs that are \emph{aimed} at metric regions rather than merely
adjacent to known designs.

\textbf{(3) Coordinated exploration operators.}
Local search plateaus on this design space because high-fitness
families sit a median of 32--37 coordinated parameter changes away from the
corpus elites: widening the front end pays off only when ROB, issue
queue, and register-file resources scale together, so single-parameter
mutation receives \emph{negative} reinforcement at every intermediate
step (Fig.~\ref{fig:barrier}). The pool
therefore includes three operators that move many parameters at once:
uniform \emph{elite crossover} between top-ranked measured designs;
\emph{path-relinking}, adopting a random 25--75\% of the differing
parameters from a donor (a measured design 70\% of the time, a random
configuration otherwise); and \emph{balanced width-group scaling},
which shifts all eight superscalar-width parameters together, usually
rescales the back-end structures in the same direction, and re-imposes
the architectural constraints. These operators encode
workload-agnostic pipeline-balance structure rather than any reference
design. The pool composition is fixed: 25\% local mutations ($\leq$6
parameters) of the top-128 measured designs, for exploitation near the
elites; 15\% each for inverse mutations, elite crossover,
path-relinking, and group scaling; and 15\% uniform-random proposals.

\textbf{(4) Distance-aware batch selection.}
The acquisition ranks candidates by two scores: a conservative score
$(\mu - 0.5\sigma)\cdot p_{\mathrm{feas}}$ that exploits, and an
exploratory score
$(\mu + \sigma)\cdot p_{\mathrm{feas}} + \lambda\, d(x)\, f^{\ast}$
that rewards distance from measured data, with $\mu, \sigma$ the
ensemble mean and spread of predicted fitness, $p_{\mathrm{feas}}$ the
predicted completion probability, $d(x)$ the normalized Hamming
distance from $x$ to $\mathcal{D}_r$, $f^{\ast}$ the best measured
fitness, and $\lambda{=}0.15$. The
distance term compensates for a failure mode of bootstrapped tree
ensembles: outside the training support all members saturate toward
the same prediction, so the spread $\sigma$ \emph{collapses} exactly
where uncertainty is largest. Each batch reserves minimum slots per
proposal family (four each for inverse, crossover, relinking, and
group-scaling proposals, two random) and fills the remainder with the
top conservative, top exploratory, and maximally diverse candidates.

\textbf{Online fine-tuning of the inverse model.}
Every four rounds, the loop fine-tunes the inverse model on accumulated
online measurements using a supervised update distinct from the
reward-guided objective of \S\ref{sec:method:rl}, mixes in stratified
corpus replay (the top-128 corpus designs by fitness plus a random
sample, 256 configurations) to prevent forgetting, and retains the
metric-normalization constants so targets beyond the corpus range
simply normalize past $1.0$. Fine-tuning enables the inverse model
to contribute inside the loop: proposals track the advancing frontier,
and \S\ref{sec:eval:search} shows it delivers the best or tied-best
median at every budget checkpoint.

\section{Evaluation}
\label{sec:eval}

\begin{figure*}[t]\centering
  \includegraphics[width=\textwidth]{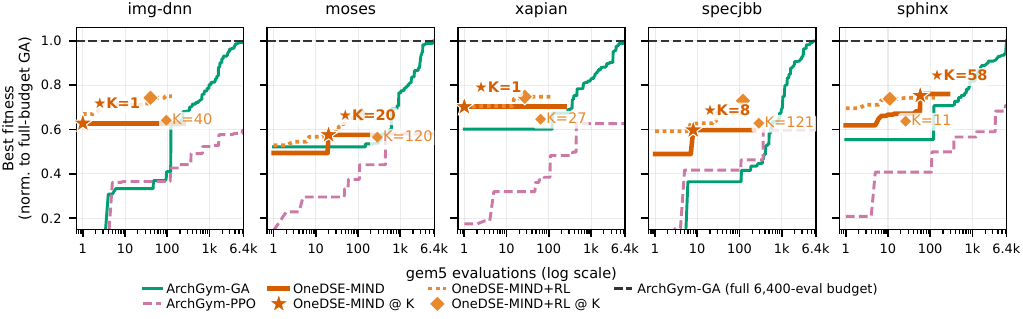}
  \vspace{-5mm}
  \caption{\NAME prediction vs.\ ArchGym-GA/PPO (best fitness
  normalized to the full-budget GA optimum; evaluations, log scale).
  $\bigstar K$ and $\blacklozenge K$ validations to reach 99\% of \NAME and \NAMERL eventual best respectively.}
  \label{fig:prediction}
  \vspace{-7mm}
\end{figure*}

Our evaluation answers two questions: how does \NAME's
metric-conditioned \emph{prediction} compare against the ArchGym
reference search algorithms~\cite{krishnan2023archgym}
(\S\ref{sec:eval:prediction}), and how close to a full-budget search
optimum does \NAMESEARCH, our hybrid of \NAME prediction and
surrogate-guided search, get with an order of magnitude fewer
simulations (\S\ref{sec:eval:search})? The results are three-fold.
First, with as few as $K{=}1$--$58$ gem5 validations, \NAME prediction
reaches design quality that ArchGym-GA needs $11\times$--$357\times$ as
many evaluations to match (median $68\times$) and that ArchGym-PPO
never matches on four of five workloads
(Fig.~\ref{fig:prediction}). Second, at 512 online evaluations
\NAMESEARCH reaches a geometric-mean $0.98\times$ the full
6{,}400-evaluation ArchGym-GA optimum, exceeding it outright on sphinx
and matching it on moses, while the strongest budget-matched
baseline, SMAC, reaches $0.83\times$ (Fig.~\ref{fig:convergence},
Table~\ref{tab:main}). Third, the designs
\NAMESEARCH finds are new: none revisits a configuration ArchGym ever
evaluated, and they cross coordinated multi-parameter barriers that
defeat mutation-based search (\S\ref{sec:eval:analysis}).

\subsection{Methodology}
\label{sec:eval:method}

\textbf{Setup.}
Every candidate is evaluated in gem5 v25.1
syscall-emulation mode (x86-64) with McPAT area models; an evaluation
warms caches and predictors for $10$\,M instructions, then measures the next $10$\,M, and takes ${\approx}212$\,s. The inverse model is trained offline on the random corpus, consuming the trace of that same post-warm-up window, so model input and measured IPC span identical instructions. 
The fitness is reported in millions (M).

\textbf{Baselines.}
Two groups of baselines separate what each method is given. The \emph{cold-start} group receives no offline corpus: ArchGym's genetic algorithm
(\emph{ArchGym-GA}, population~64)
and proximal policy optimization (\emph{ArchGym-PPO}, 32 environments), both with ArchGym's released
configurations, define the full-budget references at 6{,}400 evaluations per workload. Three additional
full-budget GA seeds on moses and sphinx characterize reference
variance, and random search at 512 evaluations provides an uninformed
floor. The \emph{warm-started} group receives the same offline corpus as \NAMESEARCH{}
and runs the same 512-evaluation online budget: ArchGym-GA (population~$64$;
re-running at population~$16$ moves the median by ${<}1\%$), GP-EI
Bayesian optimization, and SMAC.

\textbf{Budget accounting.}
\NAMESEARCH and every warm-started baseline consume the same offline
corpus: the 1{,}000 random samples together with the first 1{,}024 ArchGym evaluations. We report quality at equal
\emph{online} budget (512 new evaluations) and against the cold-start GA reference truncated at the equal \emph{total} budget (corpus\,$+$\,512
$=$ 2{,}536 evaluations). Sphinx's corpus is random-only, for 1{,}512
total evaluations.

\textbf{Implementation.}
The surrogate ensemble uses 300-tree gradient-boosted regressors
(learning rate $0.05$, 31 leaves, $0.85$ row and column subsampling),
refit from scratch each round, alongside a 200-tree feasibility
classifier with balanced class weights. Group scaling draws its width
delta from $\{-1,+1,+2\}$ with probabilities $\{0.2,0.5,0.3\}$ and
applies back-end rescaling with probability $0.7$ and a
one-to-two-parameter mutation with probability $0.5$. The elite set
comprises the top-128 measured designs, and each batch fills its
post-quota remainder half with conservative candidates, a quarter
with exploratory ones, and the rest with maximally diverse ones. Online fine-tuning of the inverse model runs one epoch per refresh at
learning rate $10^{-4}$, with 24 trace chunks per configuration and
the 256-configuration replay, taking about 16 minutes per refresh and
three refreshes per run. All learning runs on a single H100 GPU;
training \NAME{} is a one-time investment of about $60$\,h (10
epochs), amortized across every prediction target and search seed,
and reward-guided fine-tuning adds 77--89\,s per workload
(\S\ref{sec:method:rl}). All loop hyperparameters were fixed before the multi-workload
campaign and reused unchanged across workloads and seeds.

\textbf{Seeds and statistics.}
Results are medians over 5 seeds per workload (10 on img-dnn for
\NAMESEARCH, its ablations, GP-EI, and SMAC), with IQR bands in curves
and min--max whiskers in bars; significance is assessed with
Mann--Whitney $U$ on img-dnn. The online budget of 512 evaluations runs
as 16 rounds of 32-simulation batches.

\subsection{\NAME Prediction vs.\ ArchGym Search}
\label{sec:eval:prediction}

Fig.~\ref{fig:prediction} compares \NAME's prediction against
ArchGym-GA and ArchGym-PPO as a function of gem5 evaluations. \NAME
produces candidates by \emph{inverting} target metrics: we sweep a grid
of (IPC, fitness) targets, predict one configuration per target, and
validate predictions in gem5 in descending-target order; the sweep's
400 targets decode to $221$--$263$ unique designs per workload. $K$
denotes the number of validations at which \NAME reaches 99\% of its
eventual best.

\textbf{\NAME dominates the low-budget regime.}
With $K{=}1$ (img-dnn, xapian) to $K{=}58$ (sphinx) validations, \NAME
reaches designs that ArchGym-GA needs $207$--$644$ evaluations to
match, $11\times$--$357\times$ the corresponding $K$ (median
$68\times$), and that ArchGym-PPO does not match within its
\emph{entire} 6{,}400-evaluation budget on four of five workloads (on
moses, PPO eventually edges \NAME by $0.9\%$ after 6{,}400 evaluations,
$320\times$ \NAME's $K{=}20$). The corpus and training behind these
predictions are a one-time, target-agnostic investment: the trained
model answers any metric target without further search, and the same
corpus seeds \NAMESEARCH unchanged. Because
the validations are independent, they run as parallel batches of 32:
the first batch (${\approx}3.5$ minutes) already contains the eventual
best design on four of five workloads ($K{\le}32$), sphinx needs a
second ($K{=}58$, ${\approx}7$ minutes), and even the full sweep costs
only ${\approx}25$--$31$ minutes on the same 32 workers,
versus 27--91 minutes for ArchGym-GA to reach parity and roughly
700--900 minutes for its full budget.
Reward-guided fine-tuning of the inverse model (\NAMERL; dotted curves) lifts the best validated
prediction on four of five workloads, a geomean $1.12\times$ the base
prediction at $K'{=}11$--$121$ validations.

\textbf{Prediction alone is not enough.}
Fig.~\ref{fig:prediction} also exposes the limit that motivates
\NAMESEARCH: given thousands of additional evaluations, GA eventually
overtakes static prediction on every workload, because the inverse
model cannot extrapolate far beyond its random training distribution
(prediction reaches $0.58$--$0.76$ of the full-budget GA optimum). Closing this gap without paying GA's
budget is the role of \NAMESEARCH.

\begin{figure}[t]\centering
  \includegraphics[width=\columnwidth]{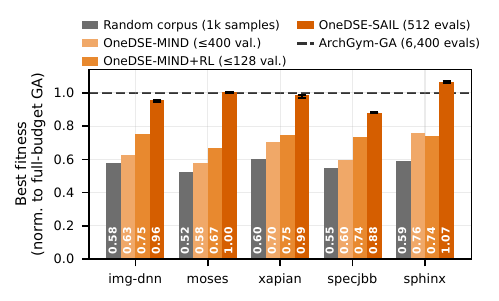}
  \caption{From corpus to prediction to search (best fitness,
  normalized to the full-budget GA optimum)}
  \label{fig:progression}
  \vspace{-5mm}
\end{figure}

\subsection{From Prediction to \NAMESEARCH}
\label{sec:eval:search}

\NAMESEARCH (\S\ref{sec:method:search}) embeds the \emph{fine-tuned}
inverse model in the measurement loop, in which the model's metric-targeted proposals
buy sample efficiency at small budgets, while the coordinated
operators carry the endpoint across the multi-parameter barriers. Fig.~\ref{fig:progression} summarizes the
progression: from a random corpus at $0.52$--$0.60$ of the full-budget
GA optimum, prediction reaches $0.58$--$0.76$, \NAMERL lifts it to
$0.67$--$0.75$, and \NAMESEARCH reaches $0.88$--$1.07$ with 512 online
evaluations.

\begin{figure*}[t]\centering
  \includegraphics[width=0.9\textwidth]{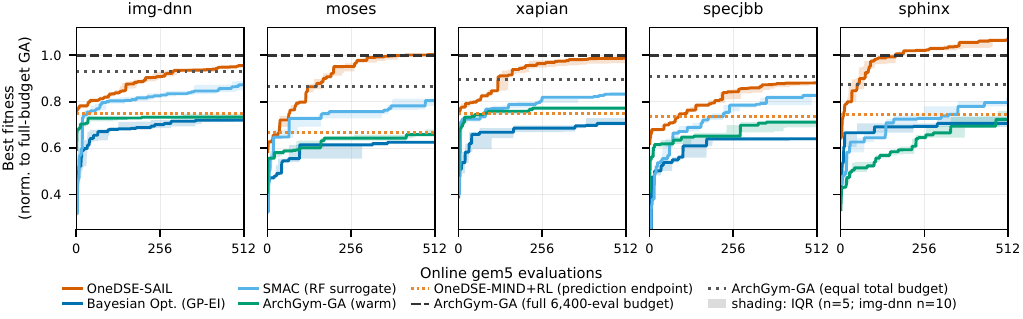}
  \caption{Equal-online-budget convergence (medians; IQR shading;
  $n{=}5$, img-dnn $n{=}10$). Dashed: full-budget GA optimum and dotted gray: equal-total-budget GA (both cold-start); dotted orange:
  \NAMERL prediction endpoint.}
  \label{fig:convergence}
  \vspace{-5mm}
\end{figure*}

\textbf{Against the cold-start references.}
Fig.~\ref{fig:convergence} and Table~\ref{tab:main} provide the main
result. At 512 online evaluations, \NAMESEARCH achieves a
geometric-mean $0.98\times$ the \emph{full} 6{,}400-evaluation
ArchGym-GA optimum, exceeding it outright on sphinx ($1.07\times$,
above all three characterized GA seeds) and matching it on moses
($1.00\times$, within the GA seed range). Charged for \emph{every}
simulation it consumes, the full offline corpus plus its 512 online
evaluations (1{,}512--2{,}536 total),
\NAMESEARCH{} still beats the cold-start equal-total-budget GA
reference by $9.1\%$ (geomean). Its median wins on four of five
workloads, its best seed on all five, and it reaches the $0.98\times$
figure with $2.8\times$ fewer simulations than the full-budget
reference. The corpus is paid once per workload, so amortized over
repeated design targets the marginal cost of each new target falls to
512 evaluations, $12.5\times$ fewer than re-running a full-budget
search.

\textbf{Against the warm-started baselines: equal information, equal
online budget.}
From the identical corpus and 512 online evaluations, the strongest
baseline, SMAC, reaches $0.83\times$ the full-budget optimum,
warm-started ArchGym-GA $0.72\times$, and GP-EI $0.68\times$
(Table~\ref{tab:main}). SMAC sits $6.1$--$25.3\%$ below
\NAMESEARCH on every workload. The advantages over SMAC, GP-EI, and
warm GA are significant (img-dnn: $p{=}9.1{\times}10^{-5}$,
$9.1{\times}10^{-5}$, and $3.3{\times}10^{-4}$; Mann--Whitney $U$,
$n{=}10$, $10$, and $5$). Cold random search trails behind ($0.793$M versus $1.405$M). The one workload where
a baseline edges \NAMESEARCH is specjbb, where warm GA leads by $3.1\%$;
\S\ref{sec:eval:analysis} traces this to simulator crashes rather
than search quality.

\begin{figure}[t]\centering
  \includegraphics[width=\columnwidth]{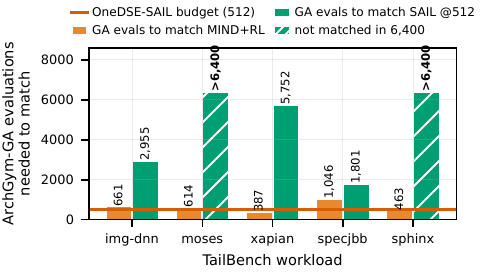}
  \caption{ArchGym-GA evaluations needed to match \NAMERL prediction and \NAMESEARCH's 512-evaluation median. Hatched: never matched in 6{,}400.}
  \label{fig:efficiency}
\end{figure}

\begin{table}[t]\centering\small
  \resizebox{\columnwidth}{!}{\begin{tabular}{@{}lrrrrrr@{}}
\toprule
& \multicolumn{4}{c}{@512 online evals} & \multicolumn{2}{c}{ArchGym-GA ref.} \\
\cmidrule(lr){2-5}\cmidrule(lr){6-7}
& \textbf{SAIL} & SMAC & GA & GP-EI & equal & full \\
\midrule
img-dnn & \textbf{1.405} & 1.282 & 1.078 & 1.060 & 1.368 & 1.470 \\
moses & \textbf{1.620} & 1.300 & 1.064 & 1.010 & 1.398 & 1.615 \\
xapian & \textbf{1.864} & 1.570 & 1.457 & 1.333 & 1.692 & 1.887 \\
specjbb & \textbf{1.352} & 1.270 & 1.093 & 0.983 & 1.394 & 1.535 \\
sphinx & \textbf{1.903} & 1.422 & 1.292 & 1.261 & 1.562 & 1.785 \\
\midrule
geomean$^{\dagger}$ & \textbf{0.98} & 0.83 & 0.72 & 0.68 & 0.89 & 1.00 \\
\bottomrule
\end{tabular}
}
  \caption{Median best fitness (M) at 512 online evaluations vs.\ the cold-start equal-total and full-budget GA references.}
  \label{tab:main}
  \vspace{-5mm}
\end{table}
\textbf{Sample efficiency.}
Fig.~\ref{fig:efficiency} inverts the comparison: ArchGym-GA needs
$1{,}801$--$5{,}752$ evaluations, $3.5\times$--$11.2\times$
\NAMESEARCH's online budget, to match \NAMESEARCH's 512-evaluation
median. On sphinx, it never does within its full 6{,}400-evaluation
budget, and on moses the canonical run falls $0.3\%$ short of ever
matching. Even the cheaper \NAMERL prediction endpoint (at most 128
validations) costs GA $387$--$1{,}046$ evaluations to match. On sphinx,
built from a random-only corpus, the advantage exceeds $4\times$ even
counting every corpus simulation against \NAMESEARCH.

\textbf{Seeding the search with the RL-fine-tuned model.}
Replacing the search's internal inverse model with its reward-fine-tuned
variant (\NAMESEARCHRL; $n{=}3$) shows that
better seeds make the loop start faster. The proposals' mean measured
fitness over the first 64 evaluations rises on four of five workloads
(up to $+9\%$ on moses), and on img-dnn the advantage persists to the
endpoint ($1.439$M median vs.\ $1.405$M). Its best seed, $1.484$M,
exceeds even the full-budget GA optimum, as does sphinx's $1.975$M, the
best single designs found by any method in this study.

\begin{figure}[t]\centering
  \includegraphics[width=\columnwidth]{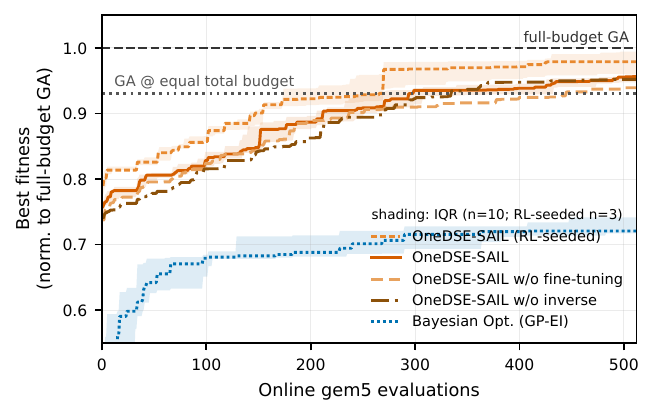}
  \caption{Ablation, img-dnn (medians; IQR; $n{=}10$, \NAMESEARCHRL
  $n{=}3$). Operators drive the endpoint (inverse removal:
  $p{=}0.62$); online fine-tuning beats frozen ($p{=}0.006$).}
  \label{fig:ablation}
  \vspace{-6mm}
\end{figure}

\textbf{Component ablation.}
Fig.~\ref{fig:ablation} ablates \NAMESEARCH on img-dnn ($n{=}10$) and
cleanly separates the two contributions. The coordinated search
operators set where the search converges: removing the inverse
model's proposals leaves the 512-evaluation median statistically
unchanged ($p{=}0.62$; a second-workload ablation on xapian confirms
the attribution, medians within seed spread, $n{=}5$). The inverse
model's in-loop contribution is real but contingent on keeping it
current: online fine-tuning significantly outperforms a frozen
inverse model at the endpoint ($p{=}0.006$). What the fine-tuned model buys is everything before the endpoint: the
best or tied-best median at every budget checkpoint, accelerated
early-budget convergence that RL-seeded initialization extends, and the standalone one-shot
prediction of \S\ref{sec:eval:prediction} at near-zero search cost. In conclusion, the fine-tuned inverse
model makes the loop sample-efficient at every early checkpoint,
RL-seeded initialization can carry that advantage to the endpoint
(img-dnn above, the best design of this study), and the coordinated
operators secure convergence across all workloads. 

\textbf{Corpus-size sensitivity.}
\NAMESEARCH{} needs far less offline data than we provide it. Cutting the
warm-start corpus $8\times$, to its first 250 designs, leaves the
512-evaluation result on img-dnn unchanged. The median stays within
2\% of the full-corpus result, and the spread across corpus sizes
($4.2\%$) is smaller than the seed-to-seed spread at any single size
($3.5$--$5.2\%$; Kruskal--Wallis over $\{250, 500, 1{,}000, 2{,}024\}$,
$p{=}0.22$; $n{=}3$ per size, 10 at the full corpus). Each smaller
corpus keeps the full corpus's mix of random and search-derived
designs, so size is the only variable. The offline campaign can
therefore be an order of magnitude smaller without penalty, moving the
cost of a new search toward its 512 online evaluations alone.

\subsection{Analysis}
\label{sec:eval:analysis}

\textbf{Crash resilience, and why specjbb favors GP-EI.}
Cycle-level simulators crash, and specjbb is the stress case:
${\approx}15\%$ of its gem5 runs fail for every method, and its dense
illegal regions crash $59.7\%$ of uniform-random proposals (versus
$4.4\%$ on sphinx), derailing any simulator-in-the-loop search that
explores broadly. Prediction-based recommendation is naturally
resilient: \NAME needs only $K{=}8$ validations on specjbb, and
\NAMESEARCH concentrates its budget on model-proposed designs. Measured
per non-crashed evaluation, \NAMESEARCH matches or exceeds warm GA on
specjbb, leading at the 256-evaluation checkpoint and still improving
at budget exhaustion. Additionally, its best seed ($1.397$M) reaches the
equal-total-budget (512) warm GA reference. 

\textbf{The discovered designs are new.}
The full-budget GA optimum on img-dnn is a balanced 8-wide front-end
design (IPC $1.38$, $9.4\,\mathrm{mm}^2$). \NAMESEARCH discovers a
distinct balanced 6-wide family (IPC $1.30$--$1.36$ at
$9.4$--$10.0\,\mathrm{mm}^2$), and in one seed independently reaches
the 8-wide family, crossing a coordinated-move barrier that defeats
single-parameter mutation. Partial width upgrades measure \emph{worse}
until ROB, register files, and issue resources scale together, which is
precisely what the coordinated operators provide. An oracle test on
img-dnn quantifies the barrier as shown in Fig~\ref{fig:barrier}, the full-budget GA elites sit a median
of 37 parameter changes from the corpus elites (single mutations reach
at most 6), and on those held-out designs the surrogate ensemble's
claimed uncertainty ($\pm 32$k fitness) understates its actual error
(${\approx}650$k) by roughly $20\times$, the uncertainty collapse that
the distance-aware acquisition term of \NAMESEARCH
compensates for.

\begin{table}[t]
\centering
\footnotesize
\resizebox{\columnwidth}{!}{%
\begin{tabular}{|l|c|c|c|}
\hline
\textbf{Method} & \textbf{DSE Time Taken (s)} & \textbf{Achieved Power (mW)} & \textbf{Achieved Latency (ns)} \\
\hline
ArchGym-GA & $1.1 \times 10^5$ & 1.23 & 0.21 \\
\hline
ArchGym-PPO & $0.7 \times 10^5$ & 1.11 & 0.32 \\
\hline
\textbf{\NAME} & \textbf{0.386} & \textbf{1.16} & \textbf{0.27} \\
\hline
\end{tabular}%
}
\caption{Extension of \NAME to DRAM memory controller DSE for a target power=1 mW and latency=0.1 ns}
\label{tab:dram_perf}
\end{table}

\begin{table}[t]
\centering
\footnotesize
\resizebox{\columnwidth}{!}{%
\begin{tabular}{|l|c|c|c|}
\hline
\textbf{Parameter} & \textbf{ArchGym-GA} & \textbf{ArchGym-PPO} & \textbf{\NAME}  \\
\hline
Page Policy & Closed & Closed & Closed  \\
\hline
Scheduler & Fifo & FrFcfsGrp & FrFcfsGrp \\
\hline
SchedulerBuffer & Shared & ReadWrite & Bankwise \\
\hline
Request Buffer Size & 1 & 4 & 16 \\
\hline
RespQueue & Reorder & Fifo & Reorder \\
\hline
Refresh Max Postponed & 4 & 8 & 4 \\
\hline
Refresh Max Pulledin & 8 & 4 & 4 \\
\hline
Arbiter & Reorder & Fifo & Reorder \\
\hline
Max Active Trans. & 1 & 1 & 1\\
\hline
\end{tabular}
}
\caption{DRAM controller parameters used by various methods, including our proposed configuration.}
\vspace{-2mm}
\label{tab:dram_param_table}
\end{table}

\begin{table}[t]
\centering
\footnotesize
\resizebox{\columnwidth}{!}{%
\begin{tabular}{|l|c|c|c|}
\hline
\textbf{Method} & \textbf{MediaBench} & \textbf{Cloud} & \textbf{PCT} \\

\hline
ArchGym-GA & 1.23 mW, 0.21 ns & 1.69 mW, 0.44 ns & 1.11 mW, 0.28 ns \\
\hline
ArchGym-PPO & 1.11 mW, 0.32 ns & 1.44 mW, 0.41 ns  & 1.09 mW, 0.28 ns \\
\hline
\textbf{\NAME} & \textbf{1.16 mW, 0.27 ns} & \textbf{1.29 mW, 0.42 ns} & \textbf{1.08 mW, 0.19 ns} \\
\hline
\end{tabular}%
}
\caption{DRAM memory controller DSE for a target power of 1 mW and target latency of 0.1 ns for diverse workloads.}
\vspace{-3mm}
\label{tab:dram_perf_many}
\end{table}

\subsection{Extension to DRAM Memory Controller DSE}
\label{subsection: dram}

To demonstrate the generalizability of our proposed framework beyond CPU microarchitecture, we extend \NAME{} to the design space exploration of the DRAM memory controller. This extension focuses on learning and predicting optimal controller parameters (such as scheduling policies and buffer sizes) while optimizing for performance and energy efficiency objectives. We utilize DRAMSys~\cite{jung2015dramsys} as both the source of representative memory traces and the baseline simulator for evaluating candidate designs. In Table~\ref{tab:dram_perf}, we present a quantitative comparison of power and latency metrics across various DSE methods for the MediaBench workload \cite{lee1997mediabench}. The target objectives for this task are a power consumption of 1 mW and an average access latency of 0.1 ns. Our results show that \NAME remains highly effective in this broader context, underscoring the framework's modularity and scalability for heterogeneous system-level DSE. Notably, \NAME achieves near-optimal design points up to $\approx3*10^5\times$ faster than ArchGym’s GA and PPO baselines. Furthermore, in Table ~\ref{tab:dram_param_table}, we present the DRAM memory controller parameter values identified by each method, highlighting the differences in design choices made by \NAME compared to baseline approaches. Additionally, in Table~\ref{tab:dram_perf_many}, we extend our exploration and demonstrate its versatility across a range of benchmarks (cloud/datacenter) sourced from \cite{srivatsankrishnan2025ossarchgym}.

\begin{table}[t]
\centering
\footnotesize
\resizebox{\columnwidth}{!}{%
\begin{tabular}{|l|c|c|c|}
\hline
\textbf{Method}
& \textbf{MACs/Cycle/mm$^2$}
& \textbf{DSE Time (s)}
& \textbf{Simulator Evaluations} \\
\hline

ArchGym-GA (p64)
& 338.63
& 376.74
& 509.2 \\
\hline

ArchGym-GA (p128)
& 349.63
& 640.67
& 1019.7 \\
\hline

\textbf{\NAME{}}
& \textbf{325.81}
& \textbf{12.59}
& \textbf{10.0} \\
\hline
\end{tabular}%
}
\caption{Extension of \NAME to FEATHER accelerator DSE averaged over 28 Llama 3.1 8B layers.}
\vspace{-6mm}
\label{tab:feather_performance}
\end{table}

\begin{table}[t]
\centering
\footnotesize
\resizebox{\columnwidth}{!}{%
\begin{tabular}{|l|c|c|c|}
\hline
\textbf{Parameter}
& \textbf{ArchGym-GA (p64)}
& \textbf{ArchGym-GA (p128)}
& \textbf{\NAME{}} \\
\hline

Dataflow
& Weight-stationary
& Input-stationary
& Weight-stationary \\
\hline

PE Array Height
& 8
& 8
& 8 \\
\hline

PE Array Width
& 256
& 256
& 64 \\
\hline

Total SRAM
& 1 MB
& 1 MB
& 0.25 MB \\
\hline

Weight precision
& 1 byte
& 1 byte
& 1 byte \\
\hline

Output precision
& 2 bytes
& 2 bytes
& 4 bytes \\
\hline

Weight load BW
& 256 bytes/cycle
& 256 bytes/cycle
& 128 bytes/cycle \\
\hline

Instruction BW
& 128 bytes/cycle
& 32 bytes/cycle
& 64 bytes/cycle \\
\hline

BIRRD drain
& 32
& 16
& 24 \\
\hline

\end{tabular}%
}
\caption{Configurations for a Llama gate/up projection ($M{=}4096$,
$K{=}4096$, $N{=}14336$)}
\vspace{-7mm}
\label{tab:feather_configurations}
\end{table}

\subsection{Extension to AI Accelerator DSE}
\label{subsection: feather}

To test transferability beyond CPU cores and memory controllers, we
apply \NAME{} to AI-accelerator DSE for GEMM workloads: the model maps
a workload and a target MACs/cycle/mm$^2$ (throughput per area) to a
candidate configuration over processing-array dimensions, on-chip
buffer capacities, operand precisions, dataflow, and memory
bandwidths. We apply \NAME{} to
FEATHER~\cite{feather}, a reconfigurable accelerator, and its MINISA
cycle-accurate simulator~\cite{minisa}, comparing against ArchGym's
GA at populations 64 and 128 (eight generations, mutation probability
0.1). On 28 held-out GEMM shapes from the Llama 3.1
8B~\cite{llama} projection layers (query/output, key/value, gate/up,
and down projections; token counts 1--4096), \NAME{} retains 97.3\%
of the ArchGym-GA population-64's MACs/cycle/mm$^2$ while being $29.9\times$ faster, and 91.8\% of the stronger population-128's
$50.9\times$ faster (Table~\ref{tab:feather_performance}). The designs are compositional rather than simply retrieved. For a gate/up projection ($M{=}4096$, $K{=}4096$, $N{=}14336$), \NAME{} selects an $8\times64$ array with 0.25 MB of SRAM, whereas both GA arms select $8\times256$ arrays with 1 MB (Table~\ref{tab:feather_configurations}). Notably, this specific configuration selected by \NAME{} is new and does not appear anywhere in the training data.

\section{Related Works}
\label{sec:related}
 
Table~\ref{tab:related_works} positions \SYS along the axes this paper
develops: search space, cost model, workload conditioning, one-shot
prediction, and sample cost.
 
\textbf{Predictors and learned simulators.} SimNet and
Tao~\cite{li2022simnet, pandey2024tao} learn the simulator itself but
are tied to one or few microarchitectures; BSSE~\cite{zheng2024bsse}
predicts metrics with Co-kNN and cuts labeling cost via
experimental-design sampling, which could substitute the random
sampling behind \NAME's corpus; TrEnDSE~\cite{wang2023transfer} and
MoDSE~\cite{wang2023modse} build ensembles with transfer learning
toward new workloads. All answer only the forward direction, on fixed
workload sets or few-shot transfer that degrades under distribution
shift (\S\ref{sec:intro}, Challenge~1). \NAME instead predicts designs
from metric targets, conditioned on the executing trace itself.
 
\textbf{Search-based DSE.} Metaheuristic
frameworks~\cite{chis2013multi, sengupta2012multi} and
ArchGym~\cite{krishnan2023archgym}, which standardizes GA, RL, ACO,
and BO agents against simulators and exposes their hyperparameter
lottery, search the parameter space with the simulator in the loop and
need thousands of evaluations. BOOM-Explorer~\cite{bai2021boom} brings
Bayesian optimization with a deep-kernel Gaussian process to BOOM
cores, and generic BO (GP-EI~\cite{jones1998efficient} along with
SMAC~\cite{smac} supplies the
strongest sample-efficient baselines. Under matched corpora and
budgets, they reach $0.68$--$0.83\times$ the full-budget reference at
512 evaluations where \NAMESEARCH{} reaches $0.98\times$
(Table~\ref{tab:related_works}; \S\ref{sec:eval:search}): the gap is
the metric-targeted inverse proposals and coordinated operators that
parameter-space acquisition does not generate.
 
\textbf{Transformers, multi-objective, and multi-agent DSE.}
Transformers have modeled parameter interactions in
DSE~\cite{xue2024multi, najafi2023deepsim, seo2025airchitect}, but
none conditions on the workload or inverts the metric map.
Multi-objective methods~\cite{palermo2005multi, abdeen2014multi,
silvano2014multi} recover Pareto fronts across PPA and compose
orthogonally with our scalar objective, as do multi-agent
formulations~\cite{krishnan2022multi, fayyazi2024arco} that decompose
the search across cooperating agents.

\section{Conclusion}
\label{section:Conclusion}

\SYSBASE{} reframes CPU DSE as inverse design: conditioned on a
workload and target metrics, it predicts designs rather than evaluating
proposed configurations. Knowing the workload increases the design
information carried by the metrics by 12--32\%, making this inversion
more tractable. \NAME{} finds strong designs in 1--58 validations that
ArchGym-GA requires 11--357$\times$ as many evaluations to match
(median $68\times$), while reward-guided fine-tuning improves prediction
quality by 12\% geomean within at most 121 validations.
\NAMESEARCH{} combines these metric-targeted proposals with coordinated
exploration and distance-aware acquisition, reaching a geometric-mean
$0.98\times$ the full 6{,}400-evaluation GA optimum with 512 online
evaluations, versus $0.83\times$ for the strongest budget-matched
baseline. It exceeds the reference on sphinx, matches it on moses, and
discovers configurations never evaluated by ArchGym. Extensions to
DRAM memory controller and the FEATHER accelerator demonstrate the applicability of \SYSBASE{} beyond CPU DSE.

\bibliographystyle{IEEEtranS}
\bibliography{ref}

\appendix
\textbf{AI usage}.
In this work, part of the experimentation code was generated by Claude code (Sonnet, Opus, and Fable), especially the plotting scripts. The key methodology ideas- \NAME and \SYSBASE-\SAILTAG{} were developed solely by the authors and AI was used only to evaluate them. AI (different generations of ChatGPT and Claude Opus and Fable) was also been used to proofread and assist the paper writing. All AI-generated prose and code were manually reviewed.

\end{document}